\documentclass[12pt]{iopart}

\usepackage{iopams}
\usepackage{setstack}
\usepackage{amsfonts}
\usepackage{amssymb}
\usepackage[dvips]{graphicx}
\usepackage{color}
\usepackage{units}
\usepackage[T1]{fontenc}
\usepackage{inputenc}
\usepackage{lmodern}
\usepackage[english]{babel}
\usepackage{epstopdf}
\usepackage[numbers,square,sort&compress]{natbib}
\usepackage[unicode=true,pdfusetitle,bookmarks=false,breaklinks=false,pdfborder={0 0 1},backref=false,colorlinks=false]{hyperref}
\usepackage{array}
\usepackage{verbatim}
\usepackage{amstext}

\newcommand{\bra}[1]{\langle #1|}
\newcommand{\ket}[1]{| #1 \rangle}

\renewcommand{\t}[1]{\mathrm{#1}}

\begin{document}

\title{The quantum Allan variance}

\author{Krzysztof Chabuda$^1$, Ian D. Leroux$^2$, Rafa{\l} Demkowicz-Dobrza{\'n}ski$^1$}

\address{$^1$ Faculty of Physics, University of Warsaw, ul. Pasteura 5, 02-093 Warszawa, Poland}
\address{$^2$ QUEST Institut, Physikalisch-Technische Bundesanstalt, 38116 Braunschweig, Germany}

\eads{demko@fuw.edu.pl}

\begin{abstract}
The instability of an atomic clock is characterized by the Allan variance,
a measure widely used to describe the noise of frequency standards.
We provide an explicit method to find the ultimate bound on the
Allan variance of an atomic clock in the most general scenario where $N$ atoms
are prepared in an arbitrarily entangled state and
arbitrary measurement and feedback are allowed,
including those exploiting coherences between succeeding interrogation steps.
While the method is rigorous and general,
it becomes numerically challenging for large $N$ and long averaging times.
\end{abstract}

\pacs{03.65.Ta, 06.30.Ft}


\maketitle

\section{Introduction} Recent years have seen spectacular improvements in the performance of optical atomic clocks \cite{Katori2011, Ludlow2015}.
This progress has been enabled by experimental techniques for trapping and manipulating ions \cite{Haffner2008} and
neutral atoms \cite{Chu2002}
previously developed for quantum information processing and simulation but now fruitfully applied to metrology.
In particular, clocks based on a single ion have demonstrated an instability of $7\times10^{-18}$ within \unit[46]{hours} of averaging~\cite{Chou2010},
while optical lattice clocks using thousands of trapped neutral atoms can now reach instabilities of $1.6\times10^{-18}$ in less than \unit[7]{hours}~\cite{Hinkley2013,Nicholson2015,AlMasoudi2015}.

In nearly all ion-based optical clocks, and a number of the latest optical lattice clocks~\cite{Nicholson2012,AlMasoudi2015}, the measurement of the accumulated atomic phase is limited by the quantum projection noise of the uncorrelated atoms in the clock.
However, unlike in optical interferometry where a phase is itself the measurand,
the purpose of an atomic clock is to continuously correct the fluctuating frequency of a classical local oscillator (LO).
Quantum projection noise therefore does not, on its own, determine achievable clock stability, which depends on a more complex interplay between phase noise, decoherence, and experimentally adjustable parameters such as the duration of the interrogation pulses applied to the atoms.
Thus, while the use of quantum entanglement to enhance clock stability has been studied theoretically for a quarter-century~\cite{Wineland1992,Wineland1994,Bollinger1996, Huelga1997, Buzek1999, Andre2004, Borregaard2013b, Giovannetti2011} and demonstrated in proof-of-concept experiments~\cite{Meyer2001,Leibfried2004, Roos2006,Louchet-Chauvet2010,Leroux2010},
it is not yet clear whether it offers significant benefits in the presence of realistic LO noise and without the artificial constraints on interrogation time imposed in demonstration experiments.

\begin{figure}
\includegraphics[width=0.75\columnwidth]{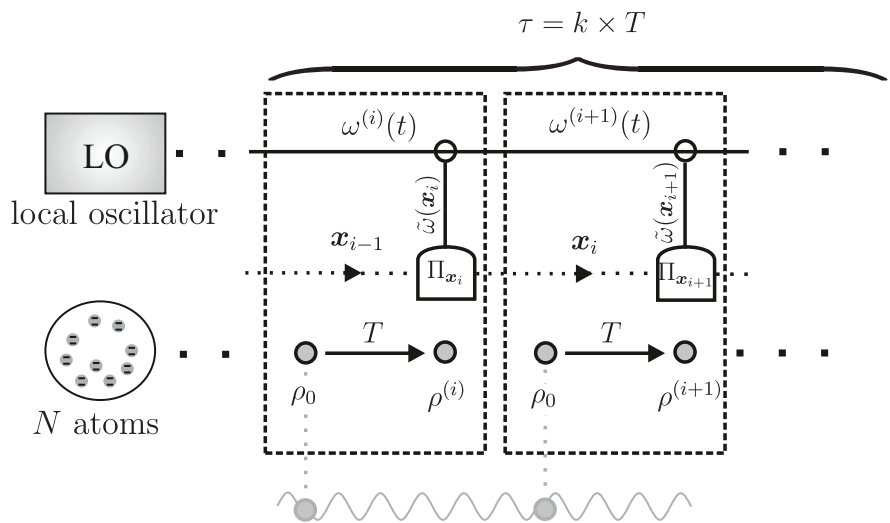}
\caption{General scheme for stabilization of a local oscillator (LO) to an atomic reference frequency $\omega_0$.
  For the $i$th interrogation, the atoms are prepared in an initial state $\rho_0$, interact with the LO whose (stabilized) frequency is $\omega^{(i)}(t)$ and evolve for a time $T$ into $\rho^{(i)}$.
  A measurement $\Pi_{\boldsymbol{x}_i}$ is applied which may depend on the outcomes of previous measurements $\boldsymbol{x}_{i-1} =  (x_1 \ldots x_{i-1})$.
  A correction signal $\tilde\omega(\boldsymbol{x}_i)$, based on all available data $\boldsymbol{x}_i$, serves to steer the LO frequency $\omega^{(i+1)}(t)$ during the next interrogation nearer to $\omega_0$.
  This cycle repeats $k$ times within an averaging interval $\tau$.
  The gray sine link at the bottom indicates that the atomic states used in different interrogation cycles may be entangled,
  so that the analysis applies to scenarios in which quantum coherences are preserved from one interrogation cycle to the next.
}
\label{fig:scheme}
\end{figure}

By comparing the performance of an existing clock or known measurement protocol to a bound valid for all possible clocks and protocols,
including those using quantum entanglement or adaptive measurement strategies,
while properly accounting for the effects of realistic noise,
one can determine whether the development and implementation of more sophisticated clock protocols is a worthwhile investment of time and resources.
We have already seen a fruitful instance of this decision-making strategy in gravitational-wave detection experiments
using quantum-enhanced optical interferometry
where it was found that the basic quantum-enhanced interferometer,
fed with coherent light and a squeezed vacuum state at the two input ports,
saturates the fundamental bounds for the precision of lossy interferometry \cite{Kolodynski2010, Knysh2010, Escher2011,Demkowicz2012} in the parameter regime relevant for gravitational-wave detection \cite{Demkowicz2013}.
This finding allows experimentalists to focus on improving technical parameters of their setup,
such as increasing laser power and reducing loss,
rather than wasting resources to prepare more sophisticated non-classical quantum states of light with no practical benefit.
This paper aims to provide analogous decision-making tools for atomic clock operation.
We present a general formula for computing, from the LO noise correlations and decoherence properties of a particular system,
a general lower bound for the Allan variance, the experimentally accessible measure of clock instability~\cite{Allan1966}.

\section{Formulation of the problem} Let $\omega(t)$ be the time-dependent frequency of the LO (laser or microwave generator)
which is to be stabilized to the reference atomic frequency $\omega_0$,
as in Fig.~\ref{fig:scheme}.
Each step of the protocol lasts for a time $T$, during which atoms prepared in a state $\rho_0$ interact with radiation from the LO.
At the $i$th step, the result $x_i$ of a measurement on the final atomic state $\rho^{(i)}$
is combined with all previous measurement results in a record  $\boldsymbol{x}_i = (x_1,\dots,x_i)$.
The LO frequency is then updated to
\begin{equation}
\label{eq:update}
\omega^{(i+1)}(t) = \omega^{(i)}(t) - \tilde{\omega}(\boldsymbol{x}_i) = \omega_{\t{LO}}(t) - \sum_{j=1}^i \tilde{\omega}(\boldsymbol{x}_j),
\end{equation}
where $\tilde{\omega}(\boldsymbol{x}_i)$ denotes the correction and
$\omega_{\t{LO}}(t)$ is the frequency of the free-running, uncorrected LO.
The figure of merit quantifying the instability of the clock is the fractional Allan variance (AVAR) \cite{Allan1966}:\begin{equation}
\label{eq:av}
\sigma^2(\tau) =
\frac{1}{2\omega_{0}^{2}\tau^{2}}\left\langle \left[\intop_{\tau}^{2\tau}\mathrm{d}t\omega\left(t\right)-\intop_{0}^{\tau}\mathrm{d}t\omega\left(t\right)\right]^{2}\right\rangle,
\end{equation}
where $\langle \cdot \rangle$ represents averaging over frequency fluctuations of the LO as well as the measurement record distribution and $\omega(t)$ is the corrected frequency of the clock: $\omega(t) = \omega^{(i)}(t)$ for $(i-1) T \leq t <  i T$.
The AVAR quantifies the expected discrepancy between two consecutive observations of the clock frequency,
each averaged over a time $\tau$.
We assume that the averaging time $\tau$ is a multiple of the interrogation time $\tau = k T$.

Given a spectrum of LO frequency fluctuations,
we want to minimize $\sigma^2(\tau)$ for a given $\tau$ and input state $\rho_0$ over
all measurement and feedback strategies.
We call the resulting minimum the \emph{quantum Allan variance} (QAVAR).
This is a horrendous task because the $2 \tau$ time period considered when calculating
$\sigma^2(\tau)$ includes a potentially large number $2 k -1$ of coupled measurement and feedback steps to optimize simultaneously.
The problem is therefore much more demanding than minimizing estimation variance in
the single-shot protocols considered in optical interferometry,
and is conceptually much closer to waveform estimation \cite{Tsang2011}.

This problem has been attacked before~\cite{Buzek1999, Andre2004, Mullan2012, Borregaard2013b, Kessler2014, Mullan2014, Macieszczak2014},
but to the best of our knowledge none of these attacks have yielded a rigorous and general bound.
In \cite{Buzek1999, Macieszczak2014} the AVAR figure of merit was replaced with an instantaneous variance which diverges for several noise processes commonly seen in experiments.
Other works such as \cite{Kessler2014} assumed to operate in the regime where correlations between $\omega(t)$ in different
interrogation steps can be neglected. 
Some analyses consider only specific protocols~\cite{Andre2004, Borregaard2013b},
or present a numerical approach that may help in improving practical interrogation schemes~\cite{Mullan2012, Mullan2014} but provides no closed formula for the fundamental bound on the Allan variance.

We consider an atomic system of $N$ two-level atoms whose evolution,
during the $i$-th step reads:
\begin{equation}
\rho^{(i)} =  U_T^{(i)}[\Lambda_T(\rho_0)],
\end{equation}
where $\Lambda_T$ is a general quantum map \cite{Nielsen2000} describing atomic decoherence processes.
Using shorthand notation $U[\rho] = U \rho U^\dagger$, the unitary operator
\begin{equation}
U^{(i)}_T =\exp\left(- \mathrm{i}  H  \int_{(i-1)T}^{iT} \t{d}t \, [\omega^{(i)}(t) - \omega_0]\right) \label{eq:Uint}
\end{equation}
describes the interrogation dynamics generated by $H= \sum_{n=0}^N n P_n$,
where $P_n$ is a projection on a subspace containing atomic states with $n$ excitations.
Note that we go to a frame rotating at the LO frequency.
For clarity we will hereafter replace $[\omega^{(i)}(t) -\omega_0]$ in the above unitary with $\omega^{(i)}(t)$,
measuring frequencies as detunings from atomic resonance,
since a constant frequency shift does not affect the AVAR. 
We choose a flat window function so the LO detuning from atomic resonance is sensed evenly over the total interrogation time $T$,
as in ideal Ramsey interrogation with two $\pi/2$ pulses of negligible duration.
A more general window function that weights the LO detuning unevenly, though possible,
generically degrades the stability of the clock through the Dick effect~\cite{Dick1987}.
We choose to ignore the Dick effect
as the main goal of the present paper is to provide fundamental stability bounds set by the intrinsic LO noise and atomic decoherence,
rather than to model in detail the effect of current technical limitations.
The ideal Dick-effect-free performance we analyze can be approached by using interleaved interrogation techniques to suppress dead time, as recently demonstrated for microwave atomic clocks~\cite{Biedermann2013} and proposed for optical lattice clocks~\cite{AlMasoudi2015}.

After the $i$-th interrogation a measurement is performed, described by a POVM $\{\Pi_{\boldsymbol{x}_i}\}$ \cite{Nielsen2000}.
This takes into account the possibility of performing generalized measurements, resulting from non-trivial interaction with ancillary systems, which are not necessarily described by projection operators.
The measurement outcomes $\{x_i\}$ occur with probabilities $p(x_i) = \t{Tr}(\rho^{(i)}\Pi_{\boldsymbol{x}_i})$.
We allow the choice of measurement operators to depend on all previous measurement results $x_j$ ($j<i$),
so that the analysis encompasses adaptive measurement protocols.
The joint probability distribution of the first $i$ measurement results reads:
\begin{equation}
\label{eq:probtotal}
p(\boldsymbol{x}_i) = \t{Tr}(\rho^{(1)} \Pi_{\boldsymbol{x}_1}) \times \dots  \times \t{Tr}(\rho^{(i)} \Pi_{\boldsymbol{x}_i}).
\end{equation}
As the feedback correction $\tilde{\omega}(\boldsymbol{x}_j)$
depends on earlier measurement results (see Eq.~\eref{eq:update}), the density matrices $\rho^{(j)}$ implicitly depend on
$\boldsymbol{x}_{j-1}$.

\section{Main result}
Given a stochastic process $\omega_{\t{LO}}(t)$ representing
frequency fluctuations of the free running LO, an initial atomic state $\rho_0$ and an interrogation time $T$,
the AVAR $\sigma^2(\tau)$ achievable in the most general stabilization scheme is lower bounded by the QAVAR $\sigma^2_Q(\tau)$:
\begin{equation}
\label{eq:result}
\sigma^2(\tau) \geq \sigma^2_Q(\tau) = \sigma^{2}_{\t{LO}}(\tau) - \frac{1}{2 \omega_0^2} \t{Tr}(\overline{\rho} L^2),
\end{equation}
where $\sigma^{2}_{\t{LO}}(\tau)$ is the AVAR of the free running LO,
\begin{equation}
\label{eq:rho}
\overline{\rho} = \left\langle \bigotimes_{i=1}^{2k-1} U_{T,\t{LO}}^{(i)}[\Lambda_T(\rho_0)] \right\rangle_{\omega_{\t{LO}}}
\end{equation}
is the tensor product of atomic states evolved at each of the interrogation steps according to the \emph{uncorrected} LO
\begin{equation}
U_{T,\t{LO}}^{(i)}=\exp{\left(- \mathrm{i} H\int_{(i-1)T}^{iT}\t{d}t\,
\omega_{\t{LO}}(t)\right)},
\end{equation}
and averaged with respect to the process $\omega_{\t{LO}}$.
The operator $L$ encapsulates information about possible feedback and measurement strategies
and is implicitly defined by
\begin{equation}
\label{eq:sld}
\overline{\rho}^\prime = \frac{1}{2}(\overline{\rho} L + L \overline{\rho}),
\end{equation}
in which
\begin{equation}
\label{eq:rhoprime}
\overline{\rho}^\prime =  \left\langle
\int_0^{\tau} \frac{\t{d}t}{\tau}\, \left[\omega_\t{LO}(t+\tau) -\omega_\t{LO}(t)\right]
\bigotimes_{i=1}^{2k-1} U_{T,\t{LO}}^{(i)}[\Lambda_T(\rho_0)]   \right\rangle_{\omega_{\t{LO}}}.
\end{equation}
A derivation of the bound is given in the \ref{app:derivation}.
A more explicit form of the QAVAR reads
\begin{equation}
\label{eq:qavexplicit}
\sigma^2_Q(\tau) = \sigma^{2}_{\t{LO}}(\tau) - \frac{1}{\omega_0^2}\sum_{rs} \frac{|\bra{\lambda_r}\overline{\rho}^\prime\ket{\lambda_s}|^2}{\lambda_r+\lambda_s},
\end{equation}
where $\ket{\lambda_r}$ and $\lambda_r$ are eigenvectors and eigenvalues of $\overline{\rho}$.
The second term in Eqs.~\eref{eq:result} and \eref{eq:qavexplicit} can be understood as a bound on the strongest
correlations that can be engineered between LO frequency fluctuations
and the corrections applied in the two averaging intervals,
given the information available from the atoms.

The QAVAR is a lower bound on the achievable AVAR, much as the Quantum Fisher Information (QFI)
is an upper bound on the classical Fisher information (FI)
that imposes a lower bound on estimation variance via the
Cram{\'e}r-Rao inequality \cite{Helstrom1976, Braunstein1994}.
Note, however, that our formulation is intrinsically Bayesian
due to the averaging with respect to the uncorrected $\omega_\t{LO}$ process, which plays the role of a prior.
In the case of the QFI it is known that there always exists a measurement for which the
FI saturates the QFI. By inspecting the derivation of the QAVAR one finds that the bound is tight
provided that one allows for experimentally implausible collective measurements on probe states at different interrogation steps.
Whether such collective measurements are indeed more powerful than adaptive strategies using separate measurements at each step is a question which we leave open. In any case, given a particular measurement and feedback strategy that approaches the QAVAR,
one is sure to have reached the ultimate optimum.

\begin{figure}
\includegraphics[width=\columnwidth]{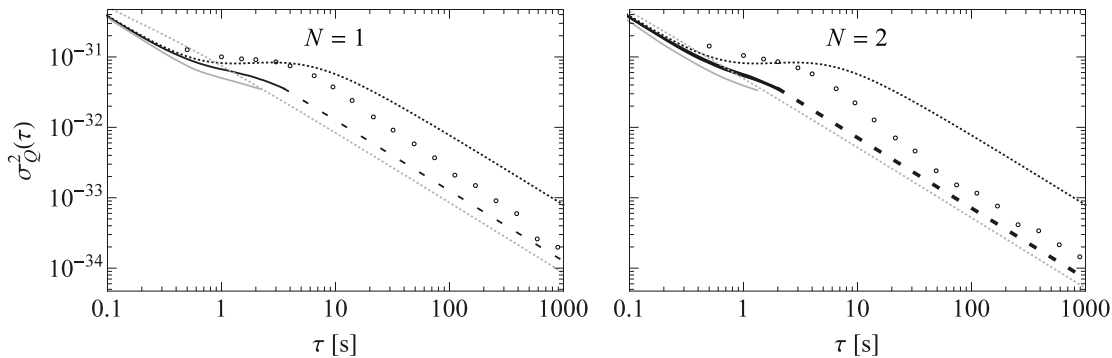}
\caption{Fractional Allan variance for atomic clock models whose LO noise (black, dotted) approximates that of Ref.~\cite{Jiang2011}.
Bounds for the AVAR are derived for clocks using a single atom ($N=1$, left) or two atoms ($N=2$, right).
In the absence of coherences between subsequent interrogations, the calculated fundamental bound (black, solid) reaches the asymptotic $1/\tau$ scaling regime, where it can be extrapolated to long times (black dashed).
In the $N=2$ case the modest gains obtainable by allowing entangled states of the two atoms lie within the thick black line.
Neglecting noise time-correlations while calculating AVAR leads to visibly looser bounds (gray, dotted) proving that white noise approximations are not justified in this setup.
Bounds for the most general scheme, allowing entanglement between the probe states used at different interrogation steps (gray, solid).
Simulations of a simple atomic clock with the same LO noise model (open circles) obey the bounds, and seem to approach them asymptotically at long times.
}
\label{fig:results}
\end{figure}
The QAVAR allows us to set aside the issue of choosing the optimal measurement and feedback and focus on the choice of
optimal probe states and interrogation time.
One can carry out brute-force minimization of the QAVAR over input probe states $\rho_0$, but a much more effective approach is
developed in \cite{Macieszczak2014, Macieszczak2013a}.
Starting with a random probe state one obtains the corresponding optimal measurement/feedback strategy described by
the operator $L$. Then, given $L$, one looks for the state performing optimally under this particular strategy,
iterating the procedure until the results converge.
For the example presented below we will assume that the initial probe states $\rho_0$, while they may involve entanglement among the atoms at a given interrogation step, are uncorrelated from interrogation to interrogation, so that the input is of the form $\rho_0^{\otimes 2k-1}$. However, we may relax this constraint and
assume that the joint state of atomic probes is arbitrary and may even involve entanglement between different interrogation steps. Though most such schemes are hardly realistic,
they cover as a special case a recent experimentally feasible
proposal where coherence is partially preserved through succeeding interrogation steps \cite{Bouyer2015}.
Such an approach, with no restrictions on independence of the atomic probes, provides the most fundamental bounds and
sets a benchmark to which realistic protocols may be compared.

These fundamental bounds also apply to protocols where different subensembles of atoms are interrogated for different times \cite{Rosenband2013, Borregard2013}.
Although we have presented our model in terms of a fixed cycle time $T$,
each cycle in the averaging time is included in our model as a separate subsystem and all these subsystems are treated formally as evolving in parallel before being measured in a collective measurement.
By allowing entanglement and joint measurements between different subsystems (time-steps), we can model interrogations that last several cycles.
For instance, consider the single-atom case 
and take two successive interrogation steps. If instead of a product state of two subsystems corresponding to two independent interrogation steps we consider an entangled state $(\ket{00} + \ket{11})/\sqrt{2}$
and let it evolve over time $T$, its evolution will be isomorphic
to the evolution of the single atom state $(\ket{0}+\ket{1})/\sqrt{2}$ evolving over interrogation time $2T$. In this way, allowing entanglement between time-steps lets us cover all schemes where effective interrogation times are multiples of some minimum cycle time $T$.

\section{Example}
\label{sec:example}
 Assume the LO  frequency fluctuations of our free-running LO are modeled by a
 combination of the Ornstein-Uhlenbeck process with white Gaussian frequency noise so that the autocorrelation function reads:
\begin{equation}
R(t) = \langle \omega_\t{LO}(t) \omega_\t{LO}(0)\rangle_{\omega_\t{LO}}  =  \alpha e^{-\gamma t} + \beta \delta(t).
\end{equation}
We choose the noise parameters to be $\alpha=\unit[2]{(rad/s)^2}$, $\beta=\unit[0.4]{(rad/s)^2 s}$, $\gamma =\unit[0.5]{s^{-1}}$, so that the
uncorrected AVAR closely resembles that of the laser system used in \cite{Jiang2011}, where $\omega_0 \approx \unit[3.25\times10^{15}]{rad/s}$, for the important region from \unit[0.1--10]{s}.
For simplicity we neglect any additional atomic decoherence effects,
which are negligible compared to LO noise in state-of-the-art atomic clocks.
In order to calculate the QAVAR we must first find the operators $\overline{\rho}$ and $\overline{\rho}^\prime$ defined in Eqs.~\eref{eq:rho} and \eref{eq:rhoprime}.
Without loss of optimality we may assume that the input state $\rho_0$ is supported on the fully symmetric subspace
of $N$ two-level atoms~\cite{Buzek1999}.
We will denote by $\ket{n}$ the symmetric state of $N$ atoms with $n$ excitations.
Let  $\ket{\boldsymbol{n}} = \ket{n_1} \otimes \dots \otimes \ket{n_K}$ be a state where $n_i$ atoms are excited at the $i$-th interrogation step. Thanks to the Gaussianity of the process we may use the cumulant expansion \cite{vankampen1981} to  obtain
an explicit formula for $\overline{\rho}$ and $\overline{\rho}^\prime$  in the $\ket{\boldsymbol{n}}$ basis (see the \ref{app:explciit} for details):
\begin{equation}
\fl \left(\overline{\rho}\right)_{\boldsymbol{n}}^{\boldsymbol{n^\prime}} = \left(\rho_0^{\otimes 2k-1}\right)^{\boldsymbol{n^\prime}}_{\boldsymbol{n}} \exp\bigg(-\frac{1}{2} \int_{0}^{(2k-1)T}\t{d}t_1 \t{d}t_2
(n^\prime_{\left\lceil\frac{t_1}{T}\right\rceil} - n_{\left\lceil\frac{t_1}{T}\right\rceil}) (n^\prime_{\left\lceil\frac{t_2}{T}\right\rceil} - n_{\left\lceil\frac{t_2}{T}\right\rceil}) R(|t_1-t_2|)\bigg),
\end{equation}
\begin{equation}
\fl \left(\overline{\rho}^\prime\right)_{\boldsymbol{n}}^{\boldsymbol{n^\prime}} =
(\overline{\rho})^{\boldsymbol{n^\prime}}_{\boldsymbol{n}} \frac{\mathrm{i}}{\tau}  \int_0^{(2k-1) T} \t{d}t_1 \int_0^\tau\t{d}t_2
(n_{\left\lceil\frac{t_1}{T}\right\rceil}-n^{\prime}_{\left\lceil\frac{t_1}{T}\right\rceil})[R(|t_1-t_2+\tau|) - R(|t_1-t_2|)],
\end{equation}
where $\left\lceil x\right\rceil$ is the ceiling function. With explicit formulas for $\overline{\rho}$ and $\overline{\rho}^\prime$, the QAVAR can be calculated using \eref{eq:qavexplicit}.

Figure~\ref{fig:results} depicts the QAVAR as a function of the averaging time for different probe states.
For each averaging time $\tau$ we look for the optimal interrogation time $T=\tau/k$ and choose the sub-multiple $k$ that minimizes the QAVAR.
This limits us to relatively short averaging times: for increasing $\tau$ the optimal $T$ stabilizes and $k$ increases linearly with $\tau$,
making the problem intractable due to the tensor product structure of $\overline{\rho}$ and $\overline{\rho}^\prime$.
Still, we expect the QAVAR to approach $\sigma_Q^2(\tau)\approx c/(\omega_0^2 \tau)$ for some constant $c$ when $\tau$ is much larger than characteristic noise correlation time scales.
Where we observe such behavior, we may extrapolate our results and estimate the best long-term clock stability obtainable with a given atomic reference and LO noise spectrum.
From the numerically optimized bounds for the current model we extrapolate rough long-term stability bounds of this form, obtaining $c \approx \unit[1.33]{rad^2\,s}$ for $N = 1$  whereas for $N = 2$  we find $\unit[0.78]{rad^2\,s}$ in the case of product states and $\unit[0.73]{rad^2\,s}$ when allowing optimally entangled states (which were almost identical to the $(\ket{00} + \ket{11})/\sqrt{2}$ Bell state).
This shows that in this case the use of entangled states provides little benefit.

At larger atom numbers,
for which the computational cost of our algorithm currently exceeds our resources,
the use of entanglement may produce greater gains.
In particular, if no additional atomic decoherence effects are taken into account, other approaches indicate the possibility of
Heisenberg scaling of Allan variance with increasing $N$ \cite{Borregaard2013b}.
Our fundamental bounds, when calculated for the corresponding noise model, must allow such scaling to be consistent with these earlier results.
That said, one must keep in mind that from some $N$ onwards even tiny levels of atomic decoherence will play a dominant role,
and it follows from general quantum metrological considerations that the scaling must revert to the standard scaling of precision in the asymptotic regime of large $N$ with at most a constant factor improvement from entanglement\cite{Escher2011, Demkowicz2012}.

We have also
plotted the QAVAR corresponding to the most general scenario where the probe states used at different interrogation steps are
optimally entangled. 
In this case we were not able to reach the $1/\tau$ regime needed to extrapolate to long-term stability.

We also plot the Allan variance observed in \unit[5000]{s} of simulated clock operation.
The modular simulator we have developed can generate LO noise traces for a range of noise processes including white, flicker, random-walk and the Ornstein-Uhlenbeck process used here, simulate the response of arbitrary references to this LO noise, and derive a frequency trace corresponding to the observed output of the frequency standard when it is stabilized to the reference according to a given feedback policy.
For the simulated clock data in Fig.~\ref{fig:results}, the atomic reference used simple Ramsey interrogation of 1 or 2 atoms in a product state, with a fixed probe time $T=\unit[0.5]{s}$.
The feedback  was derived using the integrator servo algorithm commonly used in optical frequency standards~\cite{Peik2005}.
For $N=2$, the simulated clock's stability is little worse than the bound for long averaging times, so we may conclude that a 2-ion clock operating with an LO similar to the one considered here can nearly achieve the best possible long-term stability using conventional Ramsey interrogation, and more exotic measurement strategies could provide only a minor additional improvement.

One further explanation is due in connection with \cite{Kessler2014} where Allan variance plots,
computed for clocks using a hierarchy of atomic subensembles with ever-longer interrogation times,
display an initial $1/\tau^2$
scaling only to revert to $1/\tau$ scaling once atomic decoherence becomes significant.
Even in the absence of atomic decoherence the scaling must,
for any finite $N$,
eventually revert to $1/\tau$ when there are no longer sufficient atoms to keep adding levels to the hierarchy and thus coherently monitor the LO phase for ever-longer times.
In \cite{Kessler2014} it was assumed that there were sufficient atoms available to track the laser phase up to the atomic lifetime.
In our small-$N$ example we encounter the atom-number limit too soon to observe the initial quadratic scaling.


\section{Numerical complexity of the method}
The examples provided above were limited to small values of $N$ and relatively short averaging times $\tau$. This was due to 
a rapid increase in computational cost that prevented us from going beyond the regime of $\sim1,2$ atoms and averaging times of a few seconds.
For the moment, our results are thus directly relevant to atomic clock implementations using few ions and where noise memory effects are weak enough that a few interrogation steps suffice to reach the regime where the Allan variance scales as $1/\tau$.
The computational complexity of obtaining results for larger atomic ensembles and longer noise correlation times
can be expressed as a function of two parameters, the number of atoms $N$ and the number of interrogation steps $k$ within the Allan variance averaging time $\tau$.

Although the Hilbert space describing $N$ two-level atoms is $2^N$-dimensional in general, 
for phase or frequency estimation the optimal input probe states are generally supported on the  
$N+1$-dimensional fully symmetric subspace \cite{Buzek1999,Giovannetti2006, Demkowicz2009}.
Moreover, if only LO noise is considered then the output state will be confined to this space as well, and hence the number of entires in the density matrix of 
the output state scales as $N^2$.
The same holds if, in addtition to LO noise, one allows for decoherence processes that act globally on the atoms,
such as magnetic field fluctuations that dephase all the atoms collectively.
Even for decoherence processes that violate this condition it may sometimes still be possible to describe the output density matrix in a much more efficient way than
using the full $2^N$-dimensional space. For example, if individual losses are considered, the output state can be written 
as a direct sum of states supported on fully symmetric subspaces corresponding to the different possible numbers of surviving atoms,
so that the space required to describe the output state scales as $N^3$.
Similarly in the case of individual dephasing processes,
permutation symmetry of the output density matrix (not to be confused with permutation symmetry of vectors in the Hilbert space)
admits a direct product structure where the number of parameters scales as $N^3$ \cite{jarzyna2015true}. Therefore, the scaling of 
Hilbert space dimension with the size $N$ of the atomic system,
when considering a single interrogation step,
is reasonable and without much affort
it is possible to analyze problems involving $N$ of the order of $10^2$ atoms.
A direct analysis of optical lattice clocks,
which typically use $\sim10^3$ atoms,
could be undertaken using techniques that take advantage of the fact that atoms in such setups are independent or only weakly correlated,
such as Gaussian state approximation methods.

The other parameter that dominates the numerical complexity is the averaging time $\tau$.
For averaging times much larger than the noise correlation time, we observe as expected that the optimal interrogation time tends to some value that does not depend significantly on the averaging time.
This implies that when extending $\tau$ the number $k$ of optimally tailored interrogations steps within the averaging period will scale linearly with $\tau$. Since the formalism involves a tensor product of Hilbert spaces corresponding to different interrogation steps 
the result will be an exponential scaling of the full Hilbert space with $k$ and hence with $\tau$ as well. Summing up,  the Hilbert space
that needs to be considered scales as $N^{2 k}$, and from this it is clear that the dominant source of numerical inefficiency of the algorithm is
the exponential scaling with $k$. 

Physical intuition suggest, however, that when noise correlations die out on certain characteristic time scales, it might be possible
to model the problem in a more efficient way by making use of the fact that correlations between subspaces corresponding to different interrogation 
steps will effectively be of finite range. This immediately suggests the use of 
well-developed methods from quantum information and quantum many-body physics used to describe 
systems with finite-range correlations, such as matrix product states or operators \cite{Verstraete2008, Jarzyna2013a} or
quantum Markov chains \cite{Catana2014}. Here we outline the steps needed to express our algorithm in
matrix product operator formalism. The implementation itself and its application to
realistic models will be the subject of a separate publication. 

\begin{enumerate}
\item{The first step is to choose a bond dimension for matrix
product operators, and write appropriate approximations of states
$\overline{\rho}, \overline{\rho}^\prime$ (\ref{eq:rho},\ref{eq:rhoprime}).  
One needs to investigate how large the bond
dimension must be in order to faithfully represent the state.
When noise correlations die out on a finite time scale,
a rough first choice is to take the bond
dimension to be of the order of the number of interrogation steps that
span the time scale over which noise correlations die out.}
\item{Having written the relevant states in the matrix product
operator form one plugs them into equation (\ref{eq:sld}) and finds the operator $L$,
also assumed to be in matrix product operator form. This can be
done by performing numerical minimization of a norm of the
expression $\parallel \overline{\rho}^\prime - \frac{1}{2}(\overline{\rho} L + L \overline{\rho}) \parallel$.
One can follow here a standard procedure known from matrix product
operator calculations, and perform minimization sequentially over a matrix
corresponding to a subsystem representing a given interrogation time
and proceed sequentially on different subsystems back and forth
until a satisfactory convergence is achieved.}
\item{The obtained operator $L$ is plugged into (\ref{eq:result}), which is easy as
the trace of a product of matrix product operators is directly calculable, to obtain the final bound.}
\end{enumerate}

We hope that an efficient implementation of this method will allow the treatment of averaging times long enough
to provide rigorous bounds for all typically encountered models in  present day  atomic clock setups, at least in the regime of small atom number, thanks to the fact that the increase in computational cost due to an increase of the averaging time will now only be linear instead of exponential. It is also important to reiterate that the averaging time needs to be increased only up to the point where one clearly observes 
the Allan variance taking on the $1/\tau$ scaling characteristic of white noise processes. As observed in the examples of Sec.~\ref{sec:example} this occurs for relatively short times in the models we have considered.
The prospects for analyzing practical experimental systems are therefore better than one might na{\"i}vely expect.
In order to model experimentally realistic averaging times of hours or days, it is probably sufficient to compute the QAVAR bound out to the regime of $10-100$ second averaging times, beyond which the results can be reliably extrapolated.
We indeed expect that the matrix product operator approach can enhance the numerical efficiency by the necessary one to two orders of magnitude.

\section{Conclusions}
We have provided a rigorous derivation of the fundamental bound on the AVAR in atomic clock operation, taking into account
all measurement and feedback strategies allowed by quantum theory.
We have provided explicit numerical bounds for atomic clocks using 1 or 2 atoms for a given realistic LO noise model
and have shown that, in these scenarios,
conventional Ramsey interrogation schemes perform almost as well as these bounds allow in the asymptotic regime of long averaging times.
We have also proposed an approach to deriving systematic approximations to the fundamental bound using matrix product operators,
which will allow the extension of the algorithm to more atoms and longer averaging times.
The computation of such bounds for realistic models will be a valuable guide in choosing which quantum metrological protocols or technical enhancements to pursue in order to obtain real performance enhancements in practical frequency standards.

\ack
We thank P.O.~Schmidt for helpful discussions and for a critical reading of the manuscript and K. Pachucki for sharing his
computing powers with us. This work was supported by  EU FP7
IP project SIQS co-financed by the Polish MNiSW  as well as by the Polish MNiSW ``Iuventus'' Plus
program for years 2015-2017 No. 0088/IP3/2015/73. I.D.L. acknowledges a fellowship from the Alexander von Humboldt Foundation.

\appendix
\section{Derivation of the formula for the quantum Allan variance}
\label{app:derivation}
Here we prove the main result of the paper stated in \eref{eq:result} that lower bounds the AVAR
achievable using arbitrary quantum measurement and feedback protocols by the QAVAR $\sigma^2_Q(\tau)$, given by:
\begin{equation}
\label{eq:bound}
\sigma^2(\tau) \geq \sigma^2_Q(\tau) = \sigma^{2}_{\t{LO}}(\tau) - \frac{1}{2  \omega_0^2} \t{Tr}(\overline{\rho} L^2).
\end{equation}
Let us  assume that $\tau = k T$, so that there will be $K=2k-1$ measurement/feedback steps within the time period $(0,2 \tau)$
considered when calculating the AVAR. Let $\Pi_{\boldsymbol{x}_i}$, $\tilde{\omega}(\boldsymbol{x}_i)$
describe a general measurement and feedback strategy at the $i$-th step of the protocol, where $\boldsymbol{x}_i = (x_1,\dots,x_i)$.
By definition \eref{eq:av} the AVAR reads:
\begin{equation}
\label{eq:av1}
\sigma^2(\tau) = \frac{1}{2 \omega_0^2 \tau^2} \left\langle \int \t{d}\boldsymbol{x}_K p(\boldsymbol{x}_K)
\left[\int_\tau^{2\tau}\t{d}t \omega(t) - \int_0^{\tau}\t{d}t \omega(t) \right]^2
\right \rangle_{\omega_{\t{LO}}},
\end{equation}
where
\begin{eqnarray}
p(\boldsymbol{x}_K)= \t{Tr}\left(\bigotimes_{i=1}^K \Pi_{\boldsymbol{x}_i} \rho^{(i)}\right),& \nonumber
\quad \rho^{(i)}= U_T^{(i)}[\Lambda_T(\rho_0)], \\
U_T^{(i)}= \exp\left(- \mathrm{i}  H  \int_{(i-1)T}^{iT} \t{d}t \, \omega(t)\right),&
\quad \omega(t) = \omega_\t{LO}(t) - \sum_{i=1}^{\left\lfloor\frac{t}{T}\right\rfloor} \tilde{\omega}(\boldsymbol{x}_i),
\end{eqnarray}
and $\left\lfloor x\right\rfloor$ is the floor function.
$\Lambda_T$ describes all the decoherence effects that affect the atoms during the interrogation whereas $H$
is responsible for the unitary sensing part.
Their actual form is irrelevant for the proof.
We rewrite \eref{eq:av1} as:
\begin{equation}
\fl \sigma^2(\tau) = \frac{1}{2 \omega_0^2 \tau^2} \left\langle \int \t{d}\boldsymbol{x}_K p(\boldsymbol{x}_K)
\left[\left(\int_\tau^{2\tau}\t{d}t \omega_\t{LO}(t) - \int_0^{\tau}\t{d}t \omega_\t{LO}(t)\right) - \tau \tilde{\omega}^K(\boldsymbol{x}_K)  \right]^2\right\rangle_{\omega_\t{LO}},
\end{equation}
where all feedback corrections have been gathered into
\begin{equation}
\fl \tilde{\omega}^K(\boldsymbol{x}_K) = \frac{T}{\tau} \left(  \sum_{j=k}^{K}  \sum_{i=1}^j \tilde{\omega}(\boldsymbol{x}_i) -
\sum_{j=1}^{k-1} \sum_{i=1}^j \tilde{\omega}(\boldsymbol{x}_i) \right) = \frac{T}{\tau}\left(
\sum_{i=1}^{k-1} i \tilde{\omega}(\boldsymbol{x}_i) + \sum_{i=k}^K(2k-i)\tilde{\omega}(\boldsymbol{x}_i) \right).
\end{equation}
This leads to:
\begin{eqnarray}
\sigma^2(\tau) & = \frac{1}{2 \omega_0^2 \tau^2} \left\langle
\left(\int_\tau^{2\tau}\t{d}t \omega_\t{LO}(t) - \int_0^{\tau}\t{d}t \omega_\t{LO}(t)\right)^2
\right\rangle_{\omega_\t{LO}} + \\ \nonumber
  &- \frac{1}{\omega_0^2 \tau}  \left\langle  \int \t{d}\boldsymbol{x}_K p(\boldsymbol{x}_K) \tilde{\omega}^K(\boldsymbol{x}_K) \left(\int_\tau^{2\tau}\t{d}t \omega_\t{LO}(t) - \int_0^{\tau}\t{d}t \omega_\t{LO}(t)\right)  \right\rangle_{\omega_\t{LO}} + \\ \nonumber
 & + \frac{1}{2 \omega_0^2} \left\langle \int \t{d}\boldsymbol{x}_K p(\boldsymbol{x}_K) [\tilde{\omega}^K(\boldsymbol{x}_K)]^2  \right\rangle_{ \omega_\t{LO}}.
\end{eqnarray}
Note that in the first term we performed a trivial integral over  $\int \t{d}\boldsymbol{x}_K p(\boldsymbol{x}_K)$ and what we
get is the uncorrected AVAR $\sigma^2_{\t{LO}}$.
In order to deal with the other two terms we define:
\begin{eqnarray}
\widetilde{\Pi}_{\boldsymbol{x}_i} = \widetilde{U}^{(i)}_T[\Pi_{\boldsymbol{x}_i}],& \nonumber
\quad \widetilde{U}_T^{(i)}= \exp\left(- \mathrm{i}  H T \sum_{j=1}^{i-1} \tilde{\omega}(\boldsymbol{x}_j)\right), \\
\rho^{(i)}_{\t{LO}}= U_{T,\t{LO}}^{(i)}[\Lambda_T(\rho_0)],&
\quad U_{T,\t{LO}}^{(i)}= \exp\left(- \mathrm{i}  H  \int_{(i-1)T}^{iT} \t{d}t \, \omega_\t{LO}(t)\right),
\end{eqnarray}
so that $p(\boldsymbol{x}_K)$ can be written as:
\begin{equation}
p(\boldsymbol{x}_K) = \t{Tr}\left(\bigotimes_{i=1}^K \widetilde{\Pi}_{\boldsymbol{x}_i} \rho^{(i)}_{\t{LO}}\right).
\end{equation}
The advantage of this form is that all the dependence on $\boldsymbol{x}_i$ is now in $\widetilde{\Pi}_{\boldsymbol{x}_i}$ whereas
 $\rho^{(i)}_{\t{LO}}$ represents evolution of the state assuming no corrections to a free-running LO had been applied.
As a result we arrive at:
\begin{equation}
\label{eq:av2}
\sigma^2(\tau)  = \sigma^2_{\t{LO}}(\tau)  - \frac{1}{\omega_0^2} \t{Tr} \left( L_1 \overline{\rho}^\prime \right)
+ \frac{1}{2 \omega_0^2}   \t{Tr} \left( L_2 \overline{\rho} \right),
\end{equation}
where
\begin{eqnarray}
\overline{\rho} = \left\langle \bigotimes_{i=1}^{K} \rho^{(i)}_{\t{LO}}\right\rangle_{\omega_\t{LO}}, \nonumber
\quad \overline{\rho}^\prime =  \left\langle
\int_0^{\tau} \frac{\t{d}t}{\tau}\, \left[\omega_\t{LO}(t+\tau) -\omega_\t{LO}(t)\right]
\bigotimes_{i=1}^{K} \rho^{(i)}_{\t{LO}}  \right\rangle_{\omega_\t{LO}}, \\
 L_p = \int \t{d}\boldsymbol{x}_K [\tilde{\omega}^K(\boldsymbol{x}_K)]^p \bigotimes_{i=1}^K \widetilde{\Pi}_{\boldsymbol{x}_i}.
\end{eqnarray}
Minimization of $\sigma^2(\tau)$ in \eref{eq:av2} over measurement/feedback procedures, which amounts to
optimization over $L_1$, $L_2$ operators, has the same formal structure as a standard quantum Bayesian estimation problem with
a quadratic cost function \cite{Helstrom1976,Macieszczak2014}. Using the same arguments as in the standard Bayesian problem one
first proves that standard projective measurements, where $\widetilde{\Pi}_{\boldsymbol{x}_i}\widetilde{\Pi}_{\boldsymbol{x}^\prime_i} =
\delta_{\boldsymbol{x}_i,\boldsymbol{x}^\prime_i} \widetilde{\Pi}_{\boldsymbol{x}_i}$,  are sufficient to reach the optimal performance and hence:
\begin{equation}
\label{eq:L}
\fl L_p = \sum_{\boldsymbol{x}_K} [\tilde{\omega}^K(\boldsymbol{x}_K)]^p \bigotimes_{i=1}^K \widetilde{\Pi}_{\boldsymbol{x}_i} =
\sum_{\boldsymbol{x}_K}[\tilde{\omega}^K(\boldsymbol{x}_K)  \bigotimes_{i=1}^K \widetilde{\Pi}_{\boldsymbol{x}_i}]^p = L^p,
 \quad L= \sum_{\boldsymbol{x}_K} \tilde{\omega}^K(\boldsymbol{x}_K)  \bigotimes_{i=1}^K \widetilde{\Pi}_{\boldsymbol{x}_i}.
\end{equation}
Equation \eref{eq:av2} now reads:
\begin{equation}
\label{eq:av3}
\sigma^2(\tau)  = \sigma^2_{\t{LO}}(\tau)  - \frac{1}{\omega_0^2} \t{Tr} \left( L \overline{\rho}^\prime \right)
+ \frac{1}{2 \omega_0^2}   \t{Tr} \left( L^2 \overline{\rho} \right).
\end{equation}
We should now minimize the above formula over $L$. This is a hard task if we want to keep the constraints imposed on $L$ by the structure seen in \eref{eq:L}. However, we can obtain a lower bound on $\sigma^2(\tau)$ by unconstrained minimization of \eref{eq:av3} over arbitrary hermitian operators $L$, just as in the solution to the standard quantum Bayesian problem
\cite{Helstrom1976,Macieszczak2014}.
Physically, this approach is equivalent to relaxing the product
structure of measurement operators appearing in \eref{eq:L} and allowing for arbitrary collective measurements
performed on all subsystems representing different interrogation steps of the protocol.
Differentiating \eref{eq:av3} over $L$ we obtain the condition for the optimal $L$:
\begin{equation}
\label{eq:sldsupp}
\overline{\rho}^\prime = \frac{1}{2}(\overline{\rho} L + L \overline{\rho}),
\end{equation}
which when substituted into \eref{eq:av3} yields the final bound \eref{eq:bound}, where the inequality is due to our
allowance for collective measurements.

In order to obtain a more explicit form of \eref{eq:bound} we take the matrix elements of \eref{eq:sldsupp} between eigenvectors $\ket{\lambda_r}$ of
$\overline{\rho}$ and get:
\begin{equation}
\bra{\lambda_r}\overline{\rho}^\prime\ket{\lambda_s} = \frac{1}{2} \bra{\lambda_r} L \ket{\lambda_s}(\lambda_r + \lambda_s).
\end{equation}
Solving the above equation for $\bra{\lambda_r} L \ket{\lambda_s}$ and substituting into \eref{eq:bound},
we obtain the final formula for the QAVAR:
\begin{equation}
\sigma^2_Q(\tau) = \sigma^{2}_{\t{LO}}(\tau) - \frac{1}{\omega_0^2}\sum_{rs} \frac{|\bra{\lambda_r}\overline{\rho}^\prime\ket{\lambda_s}|^2}{\lambda_r+\lambda_s}.
\end{equation}

\section{Explicit calculation of $\overline{\rho}$ and $\overline{\rho}^\prime$ for the model with gaussian LO noise}
\label{app:explciit}
We assume that apart from LO noise fluctuations described by $\omega_\t{LO}(t)$ there are no other sources of decoherence affecting the atoms.
Hence:
\begin{equation}
\fl \overline{\rho} = \left\langle \bigotimes_{i=1}^K U_{T,\t{LO}}^{(i)}[\rho_0] \right\rangle_{\omega_\t{LO}} =
\left\langle\exp\left( - \mathrm{i} \int_{0}^{K T}\t{d}t \, H^{(\left\lceil\frac{t}{T}\right\rceil)}\omega_\t{LO}(t)  \right)[\rho_0^{\otimes K}] \right\rangle_{\omega_\t{LO}},
\end{equation}
where $\left\lceil x\right\rceil$ is the ceiling function and $H^{(i)} = \sum_{n=0}^N n P^{(i)}_n$ generates evolution
of the atoms during the $i$-th interrogation step.
Without loss of generality we may assume that the input state $\rho_0$ is supported on the fully symmetric subspace
of $N$ two-level atoms~\citep{Buzek1999}, as only the total number of atomic excitations matters to the phase accumulated by an atomic eigenstate during the interrogation.
We may thus take $P_n = \ket{n}\bra{n}$, where $\ket{n}$ is the fully symmetric state of $N$ atoms with $n$ excitations.
Henceforth we will denote  by $\ket{\boldsymbol{n}} = \ket{n_1} \otimes \dots \otimes \ket{n_K}$ a state where $n_i$ atoms are excited at the $i$-th interrogation step.
When written in the $\ket{\boldsymbol{n}}$ basis,  $\overline{\rho}$ reads:
 \begin{equation}
 \label{eq:rhobarsupp}
 \left(\overline{\rho}\right)_{\boldsymbol{n}}^{\boldsymbol{n^\prime}} =  \left(\rho_0^{\otimes K}\right)^{\boldsymbol{n^\prime}}_{\boldsymbol{n}}\left\langle
 \exp\left(\mathrm{i} \int_0^{KT}\t{d}t\,
 \omega_\t{LO}(t)( n_{\left\lceil\frac{t}{T}\right\rceil} - n_{\left\lceil\frac{t}{T}\right\rceil}^{\prime}) \right)
 \right\rangle_{\omega_\t{LO}}.
 \end{equation}
When $\omega_\t{LO}$ is described by a Gaussian noise process, the cumulant expansion method \cite{vankampen1981}
implies that
\begin{equation}
\label{eq:cumulantsupp}
\left\langle e^{\mathrm{i}\int\mathrm{d}t k(t) \omega_\t{LO}(t)}\right\rangle_{\omega_\t{LO}}=
\exp\left(-\frac{1}{2}\int\mathrm{d}t_{1}\int\mathrm{d}t_{2}k(t_{1}) k(t_{2}) R \left(\left|t_{1}-t_{2}\right|\right)\right),
\end{equation}
for any arbitrary real function $k(t)$, where $R(t)$ is the autocorrelation function of $\omega_\t{LO}(t)$.
Applying \eref{eq:cumulantsupp} to \eref{eq:rhobarsupp} we get
\begin{equation}
\fl \left(\overline{\rho}\right)_{\boldsymbol{n}}^{\boldsymbol{n^\prime}} = \left(\rho_0^{\otimes K}\right)^{\boldsymbol{n^\prime}}_{\boldsymbol{n}}
\exp\left(-\frac{1}{2} \int_{0}^{KT}\t{d}t_1 \t{d}t_2\, (n^\prime_{\left\lceil\frac{t_1}{T}\right\rceil} - n_{\left\lceil\frac{t_1}{T}\right\rceil}) (n^\prime_{\left\lceil\frac{t_2}{T}\right\rceil} - n_{\left\lceil\frac{t_2}{T}\right\rceil}) R(|t_1-t_2|) \right).
\end{equation}
Analogously, using:
\begin{eqnarray}
\left\langle  \exp\left(\mathrm{i}\int\mathrm{d}t k(t) \omega_\t{LO}(t)\right) \int \t{d} t^\prime \omega_\t{LO}(t^\prime) k^\prime(t^\prime) \right\rangle_{\omega_\t{LO}}= \nonumber \\
\frac{d}{d\xi} \left.\left\langle \exp\left(\mathrm{i}\int\mathrm{d}t \left[k(t) - \xi \mathrm{i} k^\prime(t)\right]\omega_\t{LO}(t)\right)\right\rangle_{\omega_\t{LO}}\right|_{\xi=0}
\end{eqnarray}
we can obtain an explicit formula for $\overline{\rho}^\prime$:
\begin{equation}
\fl \left(\overline{\rho}^\prime\right)_{\boldsymbol{n}}^{\boldsymbol{n^\prime}} =
(\overline{\rho})^{\boldsymbol{n^\prime}}_{\boldsymbol{n}} \frac{\mathrm{i}}{\tau}
\int_0^{K T} \t{d}t_1 \int_0^\tau\t{d}t_2 \,(n_{\left\lceil\frac{t_1}{T}\right\rceil}-n^{\prime}_{\left\lceil\frac{t_1}{T}\right\rceil})[R(|t_1-t_2-\tau|) - R(|t_1-t_2|)].
\end{equation}

\bibliographystyle{iopart-num}
\bibliography{atomic}

\end{document}